\documentclass[times, 11pt]{article}
\usepackage{latex8}
\usepackage{times}
\usepackage{graphicx}
\usepackage{amsmath,amsthm,amssymb}
\usepackage{bm}
\usepackage{multirow}
\usepackage{derivative}
\usepackage{comment}
\usepackage{xcolor}

\newtheorem{Theorem}{Theorem}

\newtheorem{Example}{Example}
  
\newtheorem{Definition}{Definition}

\newcommand{\lw}[1]{\smash{\lower1.ex\hbox{#1}}}
\newcommand{\lww}[1]{\smash{\lower2.ex\hbox{#1}}}

\newfont{\Bi}{cmbxti10}

\def\QEDmark{\hspace*{\fill}\rule[0.2ex]{0.8ex}{1.4ex}}

\begin{document}
\vspace{-12pt}
\title{\Large\bf
\textrm{On the distribution of sensitivities of symmetric Boolean functions } }
\vspace{-24pt}

\author{\normalsize
\begin{tabular}[t]{c@{\extracolsep{1em}}c@{\extracolsep{1em}}c}
 \Large{Jon T. Butler} & \Large{Tsutomu Sasao} & \Large{Shinobu Nagayama}  \\
  \\
    Department of Electrical & Department of Computer Science & Department of Computer Science\\
    and Computer Engineering & and Electronics & and Network Eng.\\
    Naval Postgraduate School & Meiji University & Hiroshima City University\\
    Monterey, CA USA  93943-5121 & Kawasaki, 214-8571 JAPAN  &Hiroshima, 731-3194  JAPAN\\
    jon\_butler@msn.com & sasao@ieee.org & s\_naga@hiroshima-cu.ac.jp\\
\end{tabular}
}
\vspace{-12pt}

\maketitle
\begin{abstract} {A Boolean function $f({\vec x})$ is sensitive to bit $x_i$ if there is at least one input vector $\vec x$ and one bit $x_i$ in $\vec x$, such that changing $x_i$ changes $f$. A function has sensitivity $s$ if among all input vectors, the largest number of bits to which $f$ is sensitive is $s$. We count the $n$-variable symmetric Boolean functions that have maximum sensitivity. We show that most such functions have the largest possible sensitivity, $n$.  This suggests sensitivity is limited as a complexity measure for symmetric Boolean functions.\ \ \ \ \ \ \ \ \ }  
\end{abstract}
\newline
\vspace{-6pt}
{\bf Index Terms:}  Sensitivity, Boolean function complexity, symmetric functions.

\section{Introduction}

Sensitivity was introduced by Cook and Dwork \cite{coo82} in 1982 as a complexity measure for Boolean functions.  Since then, many papers have been written on the subject.  For example, Tur\'an, in 1984, showed that the sensitivity of all {\em non-trivial} symmetric Boolean functions is bounded below by $\lceil (n+1)/2 \rceil$, where $n$ is the number of variables.  Since no Boolean function has sensitivity greater than $n$, the sensitivities for symmetric Boolean functions span the range from $\lceil (n+1)/2 \rceil$ to $n$.

Sensitivity is one of a number of complexity measures of Boolean functions. Sensitivity has remained an enigma for about 30 years because it has resisted codification that has occurred among the other complexity measures.   Specifically, Hatami, Kulkarni, and Pankratov \cite{hat11}'s Theorem 2.7 shows seven complexity measures all of which are related by the polynomial relation.  Conspicuous by its absence is sensitivity.  In 1992, Nisan and Szegedy \cite{nis92} posed the {\bf Sensitivity Conjecture}, which posits that sensitivity is related to all seven.  In 2019, Huang \cite{hua19} solved the Sensitivity Conjecture in a remarkable two page proof.

\section{Notation}

\begin{Definition}
A Boolean function $f:\{0,1\}^n \rightarrow \{0,1\}$ whose function value is unchanged by any permutation of the input variables is {\bf symmetric}.  
\end{Definition}

\begin{Example}
The AND function $f_1 = x_1x_2$ is a 2-variable symmetric Boolean function. 
\end{Example}

\begin{Definition}
\label{sensitivity1}
Let $f\{0,1\}^n \rightarrow \{0,1\}$ be a Boolean function, and let $\vec x \in \{0,1\}^n$ be an $n$-bit {\bf input vector}. Input vector $\vec y$ is a {\bf neighbor} of $\vec x$ if $\vec y$ differs by exactly one bit from $\vec x$. $s(f,\vec x)$, the {\bf sensitivity}
of $\vec x$ in $f$, is the number of neighbors $\vec y$ of $\vec x$, such that $f(\vec x) \ne f(\vec y)$.   The {\bf sensitivity} of a function $f$, $s(f)$, is the maximum of $s(f,\vec x)$ across all choices of $\vec x$. 
\end{Definition}

\begin{Example}
Consider the 2-variable AND function, $f_1$. For the input vector, $x_1x_2 = 11$, changing either 1-bit changes the function value, so the sensitivity of this input vector is 2.  Since this is the maximum sensitivity among all input vectors, $f_1$ has sensitivity 2.
\end{Example}

\section{Relation Between the Sensitivity of a Symmetric Boolean Function and Its Composition}

In this section, we show how each symmetric Boolean function's sensitivity is determined by the function's distribution of function values, where the latter is determined by the composition associated with the function's output vector.

\vspace{-18pt}

\begin{table}[htb]
\caption{Compact Truth Tables, Compositions, and Sensitivities For All 3-Variable Symmetric Boolean Functions.}
\label{tb3}
\begin{center}
\resizebox{5.5 cm}{!}{ 

\begin{tabular}{|| c  c   c  c ||    r | c||}
\hline\hline
\multicolumn{4}{||c||}{Compact}    & \multicolumn{1}{c||}{Compo-} & \multicolumn{1}{c||}{Sensi-}\\
\multicolumn{4}{||c||}{Truth Tables}& \multicolumn{1}{c||}{sitions}& \multicolumn{1}{c||}{tiviies}\\
\hline\hline

1   &   1   &   1   &   1   &       4  & 0\\
1   &   1   &   1   &   0   &      3+1 & 3\\
1   &   1   &   0   &   1   &    2+1+1 & 3\\
1   &   1   &   0   &   0   &      2+2 & 2\\
1   &   0   &   1   &   1   &    1+1+2 & 3\\
1   &   0   &   1   &   0   &  1+1+1+1 & 3\\
1   &   0   &   0   &   1   &    1+2+1 & 3\\
1   &   0   &   0   &   0   &      1+3 & 3\\
0   &   1   &   1   &   1   &      1+3 & 3\\
0   &   1   &   1   &   0   &    1+2+1 & 3\\
0   &   1   &   0   &   1   &  1+1+1+1 & 3\\
0   &   1   &   0   &   0   &    1+1+2 & 3\\
0   &   0   &   1   &   1   &      2+2 & 2\\
0   &   0   &   1   &   0   &    2+1+1 & 3\\
0   &   0   &   0   &   1   &      3+1 & 3\\
0   &   0   &   0   &   0   &        4 & 0\\
\hline\hline
0   &   1   &   2   &   3               &\multicolumn{1}{c||}{Compo-}  & \multicolumn{1}{c||}{Sensi-}\\
\multicolumn{4}{||c||}{\# of Variables} & \multicolumn{1}{c||}{sitions}& \multicolumn{1}{c||}{tivities}\\ \multicolumn{4}{||c||}{That Are 1}      & \multicolumn{1}{c||}{      } & \multicolumn{1}{c||}{ }     \\
\hline\hline
\end{tabular}}
\end{center}
\end{table}

\vspace{-9pt}
We show a relation between $n$-variable symmetric Boolean functions and compositions on $n + 1$.   Table \ref{tb3} shows an example for $n = 3$.  The leftmost column labeled ``Compact Truth Tables'' shows the compact truth table of each of the sixteen 3-variable symmetric Boolean functions. Here, the compact truth table is shown as a {\em horizontal} vector of function values. In the compact truth table of a 3-variable symmetric function, there are four entries representing the function's value when, from left to right, zero, one, two, and three of the three variables are 1.  We include within the set of symmetric Boolean functions, the constant 0 and constant 1 functions, which are always 0 and 1, respectively, regardless of the values of the three variable values.

The second column shows compositions of integer 4.  The parts of a composition are specified by the {\em runs} of 0's and 1's in the compact truth tables starting from the left.  So, for example, the compact truth tables, 1110, 1010, and 0101 correspond to compositions 3+1, 1+1+1+1, and 1+1+1+1, respectively.

The third column shows the sensitivities of each symmetric Boolean function. Interestingly, the functions that have maximum sensitivity, 3, correspond exactly to those functions whose composition has at least one 1.

\begin{Theorem}
\label{th1}
An  $n$-variable symmetric Boolean function $f$ has sensitivity $s=n$ iff the smallest non-zero summand in the composition of $f$ is 1.
\end{Theorem}

\noindent
{\bf Proof}
\noindent
{\bf only if}: On the contrary, assume that a smallest non-zero summand value $\sigma$ has the property $\sigma > 1$.  A change in an input value can only move to an adjacent entry of the compact truth table, at least one of which maps to the same function value.   It follows that not all input values can change the function value. It follows that the sensitivity is less than $n$, a contradiction.  

\noindent
{\bf if}:
Consider an entry, $\delta$, in a compact truth table of $f$, corresponding to a summand value of 1.  Then, all edges from that entry are also incident to an entry that maps to a function value different from that of $\delta$. Changing {\em any} entry corresponding to $\delta$ changes the function value.   Thus, all $n$ elements are sensitive, and $s = n$.  Since this is maximum among all entries, $f$ has sensitivity $n$.

\QEDmark

The significance of this result is that it allows us to count functions where $s = n$.  

\section{Generating Function For the Number of Symmetric Boolean Functions With Maximum Sensitivity}

Let $A(z)$ be the ordinary generating function for the $n$-variable symmetric Boolean functions that have maximum sensitivity $s=n$. We use formal variable $z$ to track the number of function variables, and we let $a_n$ be the number of $n$-variable symmetric Boolean functions with sensitivity $s=n$.  For convenience, we include in the set of $n$-variable symmetric Boolean functions the two trivial functions $f = 0$ and $f = 1$.  As these are the only trivial functions among the set of $n$-variable symmetric Boolean functions, our set includes two additional functions to the set considered by Tur\'an\cite{tur84}.  Tur\'an shows that, among the set of {\em non-trivial} $n$-variable symmetric Boolean functions, the smallest sensitivity is $\lceil (n+1)/2 \rceil$.  

We have
\begin{equation*}
A(z)=a_0 + a_1z  +  a_2z^2 +  a_3z^3  +  \cdots  +  a_n z^n  + \cdots . \\
\end{equation*}
\noindent
We develop $A(z)$  in two steps. In the first step, we derive $T(z)$, the generating function for {\em all} symmetric $n$-variable Boolean functions.  That is,

\begin{equation}
\label{e1}
T(z)\! =\! \frac{4z}{1-2z}\!=\! 4z + 8z^2  + 16 z^3 + \cdots + 2^{n+1} z^n + \cdots ,\\
\end{equation}

\noindent
wherever the power series converges.  This specifies that there are four 1-variable, eight 2-variable, 16 3-variable, and $2^{n+1}$ $n$-variable symmetric Boolean functions.

Recall from the previous section, that the $n$-variable symmetric Boolean functions with sensitivity $n$ correpond to those functions whose composition have at least one 1.   Let $N_{\hat{1}}(z)$ be the generating function for the number of symmetric Boolean functions with {\it no} 1's in their composition.  Then, the generating function for the number of symmetric Boolean functions with at least one 1 in their composition is $A(z)= T(z) - N_{\hat{1}}(z)$.

In the second step, we use the recursive method proposed by Chinn and Heubach \cite{chi03} for the generation of
\begin{table}[htb]
\label{tb5}
\begin{center}
\begin{tabular}{p{0.07\textwidth}p{0.8\textwidth}}
$C(n)$          &\!\!\!\!\!\! the number of compositions of $n$, and\\
$C(n,\hat{1})$  &\!\!\!\!\!\! the number of compositions of $n$ with no 1's.\\
\end{tabular}
\end{center}
\vspace{-10pt}
\end{table}

From Theorem 1 of \cite{chi03}
\begin{equation}
\label{eq3}
C(n,\hat{1}) = 2\cdot C(n-1,\hat{1}) - C(n-1,\hat{1}) + C(n-2,\hat{1}), \\
\end{equation}
\noindent
for $n \geq 2$.  Here, the term $2\cdot C(n-1,\hat{1})$ recursively counts compositions of $n$ that are formed by 1) adding a 1 to the last summand of compositions of $n-1$ or by 2) appending a single 1.  However, this fails for two reasons.  First, this does not count compositions on $n$ that end in a single 2.  The second term corrects this.  Second, we must subtract the compositions of $n-1$ that ended in 2, since they would now end in 1 from the recursive process.

(\ref{eq3}) becomes
\begin{equation}
\label{eq4}
C(n,\hat{1}) =C(n-1,\hat{1}) + C(n-2,\hat{1}), \\
\end{equation}

\noindent
which shows that $C(n,\hat{1})$ is the sum of the two previous $C$'s. Table \ref{tb60} shows the computation of $N_{\hat{1}}(x)$, the generating function for the number of symmetric Boolean functions associated with compositions having no 1's.

\vspace{-12pt}
\begin{table}[htb]
\caption {Generating Function for Symmetric Boolean Functions Associated With Compositions Having No 1's}
\label{tb60}
\begin{center}
\begin{tabular}{p{0.09\textwidth}p{0.8\textwidth}}
$N_{\hat{1}}(z) =$   & Compositions with no 1's.\\
$2zN_{\hat{1}}(z)$   & Form compositions of $n$ from $n-1$ by adding 1 to the last summand or appending a 1.  Later, correct for appending a 1.\\
$+z^2N_{\hat{1}}(z)$ & Because we want no 1's, there is no 2 at the end of a composition of $n$.  So, we must append 2 to compositions of $n-2$.\\
$-zN_{\hat{1}}(x)$   & Remove compositions ending in +1.\\
$+2z$        & Initial conditions are not generated above.  For $n = 1$, there are two input vectors and four functions, 00, 01, 10, and 11, two of which correspond to composition 2 (the other two correspond to 1+1).\\
\end{tabular}
\end{center}
\end{table}
\vspace{-18pt}

From Table {\ref{tb60}}, the generating function, $N_{\hat{1}}(z)$, for compositions with no 1's is

\begin{equation}
\label{e3}
N_{\hat{1}}(z)\!=\!\frac{2z}{1 - z -z^2}.
\end{equation}

\noindent
Therefore, the generating function for the number of compositions that have at least one 1 is
\begin{equation}
\label{e3333}
A(z)  = \frac{4z}{1-2z} - \frac{2z}{1 - z -z^2},
\end{equation}
\noindent
which is the generating function for the number of $n$-variable symmetric binary functions that have sensitivity $n$.

\section{Asymptotic Approximation For the Number of Symmetric Boolean Functions With Maximum Sensitivity}

In this section, we derive a simple approximation to the number of symmetric Boolean functions with maximum sensitivity as a function of the number of variables $n$.
\begin{Definition}
\label{d3}
\begin{equation}
f_n \thicksim g_n, \ \ \ \ i\!f\  \lim_{n \rightarrow \infty}\frac{f_n}{g_n} = 1.
\end{equation}
\end{Definition}
\noindent
Here, $g_n$ is an exact expression, and $f_n$ is an approximation. We use the following result by Bender \cite{ben74}.

\begin{Theorem} \cite{ben74} Suppose $A(z) = \sum_{n \geq 0} a_n z^n$ is analytic near $0$ and has only algebraic singularities on its circle of convergence. Let $w$ be the maximum of the weights at
these singularities. Denote by $\alpha_k$, $\omega_k$, and $g_k$ the values of $\alpha$, $\omega$ and $g$ for those terms of the form  $(1 -z/\alpha)^{-\omega} g(z)$ of weight $w$. Then

\begin{equation}
\label{e8}
a_n - \frac{1}{n}\sum_k \frac{g_k(\alpha_k)n^{\omega_k}}{\Gamma(\omega_k)\alpha_k^n} = o(r^{-n}n^{w-1}),
\end{equation}
\noindent
where $r=|\alpha_k|$, the radius of convergence of $A(z)$, and $\Gamma(s)$ is the gamma function.
\end{Theorem}

Recall that

\begin{equation}
\label{e91}
A(z) = \frac{4z}{1-2z} - \frac{2z}{1-z-z^2} .
\end{equation} For this generating function, the circle of convergence has radius, $\alpha_1 = (1/2)$, and so the term $\frac{4z}{1-2z}$ in (\ref{e91}) determines the asymptotic values.  We have $k = 1$, $\alpha_1 = 1/2$, $\omega_1 = 4\cdot (1/2) = 2$ and $r = |\alpha_1| = (1/2)$. With these values, (\ref{e8})
becomes

\begin{equation}
\label{e19}
a_n - \frac{1}{n}    \frac {2 \cdot n^1}  { {\Gamma(1)}     (1/2)^n        }   =    o((1/2)^{-n}n^{1-1}),
\end{equation}
\begin{equation}
  {a_n \thicksim 2^{n+1}}.
\end{equation}

Since $2^{n+1}$ is the total number of $n$-variable symmetric Boolean functions, this shows that the number of functions with sensitivity $s=n$, approaches the total number of functions as $n$ approaches $\infty$.

\section{Concluding Remarks}

A complexity measure for Boolean functions that maps all functions to the same value is useless. In this paper, we show that most symmetric Boolean functions map to sensitivity $n$, the maximum, suggesting that sensitivity is limited as a complexity measure for symmetric Boolean functions.

\section*{Acknowledgments}
This research is partly supported by three Grants-in-Aid from the Japan Society for
the Promotion of Science (JSPS) for Scientific Research, including JSPS KAKENHI Grant (C), No. 19K11881, 2022.

\end{document}